\documentclass[aps,prd,showpacs,nobibnotes,twocolumn,preprintnumbers,
nobalancelastpage,superscriptaddress]{revtex4-1}

\usepackage{natbib}
\setcitestyle{numbers}
\usepackage{amssymb} 
\usepackage{amsmath} 
\usepackage{graphicx}
\usepackage{latexsym} 
\usepackage{dcolumn}
\usepackage{bm}
\usepackage{color}
\input{epsf} 
\newcommand{\be}{\begin{equation}}
\newcommand{\ee}{\end{equation}} 
\newcommand{\ba}{\begin{eqnarray}}
\newcommand{\ea}{\end{eqnarray}} 

\newcommand{\bfi}{\begin{figure} \epsfxsize=8cm \epsffile}
\newcommand{\bfig}{\begin{figure*} \epsfxsize=15cm \epsffile}
\newcommand{\efi}{\end{figure}} 
\newcommand{\efig}{\end{figure*}}
\newcommand{\bi}{\begin{itemize}} 
\newcommand{\ei}{\end{itemize}}
\newcommand{\mpch}{h^{-1} {\rm Mpc}} 
\newcommand{\hmpc}{h {\rm Mpc}^{-1}}

\newcommand{\Msunh}{M_\odot/h}
\newcommand{\dif}{\mathrm{d}}

\newcommand{\bmx}{{\mathbf{x}}}
\newcommand{\bmr}{{\mathbf{r}}}
\newcommand{\bmk}{{\mathbf{k}}}
\newcommand{\bmv}{{\mathbf{v}}}
\newcommand{\bmve}{{\mathbf{v_E}}}
\newcommand{\bmvb}{{\mathbf{v_B}}}
\newcommand{\bmvstar}{{\mathbf{v_*}}}

\newcommand{\bibmnras}{Mon. Not. R. Astron. Soc.}
\newcommand{\bibapj}{Astrophys. J.}

\newcommand{\bibaap}{Astron. Astrophys.}
\newcommand{\bibprd}{Phys. Rev. D}
\newcommand{\bibprl}{Phys. Rev. Lett.}
\newcommand{\bibphysrep}{Phys. Rep.}

\newcommand{\bibnat}{Nature}
\newcommand{\bibjcap}{J. Cosmol. Astropart. Phys.}

\begin{document}

\title{Kriging interpolating cosmic velocity field}

\author{Yu Yu}
\email{yuyu22@shao.ac.cn}
\affiliation{Key laboratory for research in galaxies and cosmology,
Shanghai Astronomical Observatory, Chinese Academy of Science, 80
Nandan Road, Shanghai, China, 200030}

\author{Jun Zhang}
\affiliation{Center for Astronomy and Astrophysics, Department of Physics and Astronomy,
Shanghai Jiao Tong University, Shanghai, 200240}
\affiliation{IFSA Collaborative Innovation Center, Shanghai Jiao Tong University, Shanghai, China, 200240}

\author{Yipeng Jing}
\affiliation{Center for Astronomy and Astrophysics, Department of Physics and Astronomy,
Shanghai Jiao Tong University, Shanghai, 200240}
\affiliation{IFSA Collaborative Innovation Center, Shanghai Jiao Tong University, Shanghai, China, 200240}

\author{Pengjie Zhang}
\email{zhangpj@sjtu.edu.cn}
\affiliation{Center for Astronomy and Astrophysics, Department of Physics and Astronomy,
Shanghai Jiao Tong University, Shanghai, 200240}
\affiliation{IFSA Collaborative Innovation Center, Shanghai Jiao Tong University, Shanghai, China, 200240}
\affiliation{Key laboratory for research in galaxies and cosmology,
Shanghai Astronomical Observatory, Chinese Academy of Science, 80
Nandan Road, Shanghai, China, 200030}

\begin{abstract}
Volume-weighted statistics of large-scale peculiar velocity is preferred by peculiar velocity cosmology,
 since it is free of the uncertainties of galaxy density bias entangled in observed number density-weighted statistics.
However, measuring the volume-weighted velocity statistics from galaxy (halo/simulation particle) velocity data is challenging. 
Therefore, the exploration of velocity assignment methods with well-controlled sampling artifacts is of great importance.
For the first time, we apply the Kriging interpolation to obtain the volume-weighted velocity field.
Kriging is a minimum variance estimator.
It predicts the most likely velocity for each place based on the velocity at other places.
We test the performance of Kriging quantified by the E-mode velocity power spectrum from simulations.
Dependences on the variogram prior used in Kriging, the number $n_k$ of the nearby particles to interpolate, and the density $n_P$ of the observed sample are investigated.
First, we find that Kriging induces $1\%$ and $3\%$ systematics at $k\sim 0.1\hmpc$ when $n_P\sim 6\times 10^{-2} (\mpch)^{-3}$ and $n_P\sim 6\times 10^{-3} (\mpch)^{-3}$, respectively.
The deviation increases for decreasing $n_P$ and increasing $k$.
When $n_P\lesssim 6\times 10^{-4} (\mpch)^{-3}$, a smoothing effect dominates small scales,
 causing significant underestimation of the velocity power spectrum.
Second, increasing $n_k$ helps to recover small-scale power.
However, for $n_P\lesssim 6\times 10^{-4} (\mpch)^{-3}$ cases, the recovery is limited.
Finally, Kriging is more sensitive to the variogram prior for a lower sample density.
The most straightforward application of Kriging on the cosmic velocity field does not show obvious advantages over the nearest-particle method [Y. Zheng, P. Zhang, Y. Jing, W. Lin, and J. Pan, Phys. Rev. D 88, 103510 (2013)] and could not be directly applied to cosmology so far.
However, whether potential improvements may be achieved by more delicate versions of Kriging is worth further investigation.
\end{abstract}

\maketitle


\section{Introduction}
\label{sec:introduction}

The cosmic velocity field carries cosmological information.
Peculiar velocity probes the growth rate of the Universe.
This makes it powerful to separate gravity and dark energy models
(e.g., Refs. \cite{Kaiser87,Peacock01,Linder03,zhangpj07,Guzzo08,Jain08, wangyun08,Reyes10,limiao11,Clifton12,Reid12, Tojeiro12,Weinberg13,Joyce15,Koyama15}).
Obtaining the observed number density-weighted and mass-weighted velocity is straightforward from galaxy velocity data and simulations.
However, statistics involving the observed number density-weighted velocity suffers from uncertainties in galaxy density bias.
On the contrary, the volume-weighted velocity statistics is free of uncertainties from galaxy bias.
Thus, it is of particular importance since it is desired for the purpose of cosmology.
However, modelling of the volume-weighted velocity is nontrivial due to the sampling artifacts in velocity assignment methods
(e.g., Refs. \cite{zhengyi13,zhengyi15b,zhangpj15,zhengyi15a}).
These sampling artifacts arise from assigning velocity on grids from highly nonuniformly distributed particles/galaxies,
especially for the populations with a low number density.

These sampling artifacts in observation bias the velocity power spectrum measured from galaxy velocity data.
The same artifacts in simulation hamper the theoretical understanding of the velocity field.
A biased theoretical understanding can lead to biased cosmological constraints,
 even the velocity measurements are free from sampling artifacts like the one inferred from redshift-space distortion (RSD) measurement
\footnote{Since the velocity power spectrum in the RSD measurement is inferred indirectly and statistically from the redshift space galaxy clustering of which the measurement is unbiased, these velocity measurements themselves do not suffer from the sampling artifact \cite{zhangpj15}.}.
To proceed, two aspects could be improved.
A straightforward one is choosing a reasonably good velocity assignment method with negligible sampling artifacts.
Alternatively, understanding and modelling the sampling artifacts for a chosen assignment method could correct the systematics and improve the velocity statistics measurement.

In the literature, several methods have been adopted to estimate the volume-weighted velocity field.
The Voronoi tessellation (VT) method is a zeroth-order interpolation scheme.
One first constructs Voronoi tessellation from a set of nodes (i.e., particles/haloes) which forms the Voronoi polyhedral.
Each Voronoi polyhedral contains only one particle and shares a wall with another polyhedral.
Inside a Voronoi polyhedral, the velocity is approximated as a constant, which is just the velocity of the particle inside.
Smoothing this space-filling velocity field, one could obtain the velocity on regular grids.
The Delaunay tessellation (DT) method \cite{Schaap00} is a linear interpolation scheme.
One first constructs Delaunay tessellation, which is the dual of Voronoi tessellation and forms the Delaunay tetrahedron.
The velocity gradient inside a Delaunay tetrahedron is approximated as a constant
 and is determined by the velocity of the four vertices.
Also, the volume-weighted quantities are determined by applying a smoothing on the interpolated velocity field.
In Bernardeau and van de Weygaert's work \cite{Bernardeau96}, the above two methods were proposed and found to agree well for a considerable range of situations.
\citet{Bernardeau97} found that the velocity divergence probability distribution function (PDF) measured by the above two methods
 also successfully produces the dependence on the matter density parameter.
\citet{Pueblas09} adopted the DT method to define the velocity field,
 showing that the power spectra of velocity divergence and vorticity measured in this way are unbiased
 and have better noise properties than for standard interpolation methods that deal with the mass-weighted velocities.
However, theoretical modelling by \citet{zhangpj15} argues that none of these methods are free of sampling artifacts,
 in particular at the stringent $1\%$ level required by Stage IV dark energy projects.

In the works by \citet{zhengyi13} and \citet{Koda14}, the nearest-particle (NP) method was proposed and applied.
It is just the first step of the VT method that assigns the velocity of a regular grid point as the velocity of the nearest particle to it.
In this sense, it is the most straightforward velocity assignment method.
Without applying a smoothing, it alleviates artificial suppression of small-scale random motions.
Nevertheless, the NP method suffers from sampling artifacts, first detected by \citet{zhengyi13}.
More robust and systematic detection of the sampling artifacts from the NP method through N-body simulation is reported by \citet{zhengyi15a}.
It causes $\sim 12\%$ underestimation of the velocity power spectrum at $k=0.1\hmpc$
 for samples with mean number density $\sim 6\times10^{-4}(\mpch)^{-3}$.
An advantage of the NP method is that the sampling artifacts are relatively straightforward to understand.
Theoretical modelling of this sampling artifact helps to correct the systematics to some extent
\cite{zhangpj15,zhengyi15a}.
However, we are still not able to match the stringent requirement ($1\%$ modelling accuracy) of Stage IV dark energy projects.
Further improvements on modelling the sampling artifacts and/or better velocity assignment methods are desired.
This paper presents our tests of Kriging as a velocity assignment method.

Kriging, originated in geostatistics, is a method of interpolation
 for which the interpolated values are modelled by a Gaussian process governed by prior covariances.
It can be understood as a linear prediction or a form of Bayesian inference.
The Kriging method starts with a set of values observed and assumes a prior distribution over the whole field.
A new value can be predicted at any new spatial location, by combining the Gaussian prior
 with a Gaussian likelihood function for each of the observed values.
The resulting posterior distribution is also Gaussian, with a mean and covariance that can be simply computed from the observed values, their variance, and the kernel matrix derived from the prior.
Assuming the Gaussian velocity field, it gives the most likely velocity at the interpolated position
 based on the configuration of points with known velocity.
It is accurate when the interpolated position overlaps with one of the observed points.
The weighting derived from the Kriging method does not depend on the observed values.

It is worth noting that the Kriging method is also useful in other branches of astrophysics.  
For example, in the field of weak gravitational lensing, the presence of the point spread function (PSF) biases the galaxy shape measurement.
To correct it, one needs to interpolate the PSF form at the galaxy position based on the star images nearby.
It has been found that the Kriging method generally performs better than other methods, including polynomial interpolation, radial basis functions, and Delaunay triangulation, in the context of weak lensing \cite{Berge12}.
We are therefore strongly motivated to test the performance of Kriging on cosmic velocity fields.

This paper is organized as follows.
In Sec. \ref{sec:kriging}, we introduce the Kriging method in general cases and in the application on the cosmic velocity field.
The simulation we use is briefly described in Sec. \ref{sec:simulation}.
The statistics with which we are most concerned, the E-mode velocity power spectrum obtained from the Kriging method, is also introduced in this section.
We present the dependence of the Kriging method performance on various factors in Sec. \ref{sec:results}.
We conclude and discuss in Sec. \ref{sec:conclusion}.


\section{Kriging Interpolation}
\label{sec:kriging}

Under suitable assumptions on the prior, Kriging gives the best linear unbiased prediction. 
The method is widely used in the domain of spatial analysis and computer experiments. 
The technique is also known as the {\it Kolmogorov-Wiener prediction} \cite{Cressie88,Cressie93}.

\subsection{General idea of Kriging}
\label{sec:generalkriging}

We assume that we want to estimate the field at a given position $\bmx_*$ from $n_k$ observed data points at $\bmx_i$
 with a value of $y_i=y(\bmx_i)$.
Kriging looks for the value of the field at this position as a weighted linear combination of the nearby values at known positions,
\be
\hat{y}_*=\sum_i W_i y_i\ .
\ee
The weighting $\bm{W}$ is estimated such that the estimator is unbiased, and it minimizes the error with respect to the data according to the mean square variation.
Therefore, Kriging relies on the {\it variogram} of the data as a prior to interpolate.
The variogram is the mean-square variation of field values as a function of the separation,
\be
\gamma(\bmr)=\frac{1}{2}\langle [y(\bmx+\bmr)-y(\bmx)]^2 \rangle \ ,
\ee
in which the ensemble average is over all spatial positions.
Usually one assumes that the variogram only depends on the distance under the assumption of isotropy.

The weighting $\bm{W}$ is solved from the following Kriging system:
\be
\left[\begin{array}{c} \bm{W} \\ \mu \end{array}\right]
=\left[\begin{array}{cc} \bm{\gamma}_{ij} & \bm{1} \\ \bm{1^T} & 0 \end{array}\right]^{-1}
\left[\begin{array}{c} \bm{\gamma}_{i*} \\ 1 \end{array}\right]\ .
\label{eqn:kriging}
\ee
Here, $\bm{\gamma}_{ij}$ is the variogram matrix between observed points,
 with $\gamma_{ij}=\gamma(r_{ij})=\gamma(|\bmx_i-\bmx_j|)$.
It has a dimension of $n_k\times n_k$.
Unity vector $\bm{1^T}$ is set for the unbiased condition $\sum_i W_i=1$.
$\bm{\gamma}_{i*}$ is the variogram vector between the interpolated point and the observed points.
$\mu$ is Laplacian multiplier.
We refer the readers to the Appendix for the detailed derivation and some useful properties of the Kriging method.

One could measure this variogram from the observed data set itself numerically.
For general usage, one could adopt a suitable model and fit the parameters from the data set to obtain an experimental variogram.
Gaussian, spherical, and exponential variogram models are frequently used in applications.

Usually the variogram function approaches to zero when $r\rightarrow 0$.
For some cases, field values could change significantly even for small separation.
One example is the existence of a measurement error for an observed sample.
In this case, one obtains different values for multiple observations even at the same point.
Another example in cosmology is the nonlinearly evolved velocity field.
After shell crossing, multistreams exist in cosmic flow, leading to a nonzero variogram at $r=0$.
This variogram behavior is called the {\it nugget} effect.
With the nugget effect, Kriging always tries to smooth the field to suppress uncertainties existing within a small distance.
Thus, in this case, Kriging interpolation no longer gives the observed value
 even that the position to interpolate approaches one of the observed points.

\subsection{For cosmic velocity field}
\label{sec:velocitykriging}

The cosmic velocity field at large scales ($k\sim 0.1\hmpc$) is well approximated as curl free.
Thus, the large-scale behavior is well described by the velocity divergence $\theta(\bmx)\equiv-\nabla\cdot\bmv(\bmx)/H$.  
Here, the normalization $H$ is the Hubble parameter.
The large-scale velocity is coherently evolved with the density field, which is homogenous and isotropic.
Thus, we adopt homogenous and isotropic velocity variogram in the Kriging interpolation.
This variogram only depends on the separation of the two points:
\be
\gamma_{ij}=\gamma(r_{ij})=\frac{1}{2}\langle |\bmv(\bmx_i)-\bmv(\bmx_j)|^2\rangle\ .
\label{eqn:variogram}
\ee
Averaged over the directions of ${\bmr_{ij}}$, the correlation between different spatial velocity components vanishes.
Thus, each velocity component is Kriging interpolated from the same velocity component of other points.

For general Kriging interpolation, a power-law fit variogram is sufficient to obtain a reasonable good estimation.
However, a coarse fit of the velocity variogram may lead to an intolerant error in the estimated velocity power spectrum for precision cosmology.
On the theoretical aspect, we do not have an accurate variogram model for cosmic velocity due to the nonlinear evolution.
However, we could obtain the linear velocity power spectrum/correlation function prediction, and thus the linear variogram.
Although it neglects the nonlinear effect that is difficult to model, 
 the linear prediction is a good choice as the prior for the performance test.
 
We do not include the nugget effect in this performance test for several reasons.
First, although there exist velocity measurement errors for haloes/galaxies in observation,
 we only deal with simulation particles with an accurate velocity measurement.
Second, we only need the most likely velocity on grid points predicted by Kriging instead of the full posterior distribution.
The full posterior distribution is useful in Kriging fitting, in which the Nugget is an important component.
Finally, we want to keep the property that the velocity of the grid point is assigned to the velocity of the particle passing it.
This will not suppress the random motions responsible for the fingers-of-god effect at a small scale where the particles are dense.

Note that the velocity correlation $\xi_{ij}(\bmr)\equiv\langle v_i(\bmx_1)v_j(\bmx_2)\rangle$ 
 between the $i$th velocity component at position $\bmx_1$ and $j$th component at $\bmx_2=\bmx_1+\bmr$ can be decomposed into two correlation functions $\psi_\perp$ and $\psi_\parallel$ \cite{Peebles80},
\be
 \xi_{ij}(\bmr)=\psi_\perp(r)\delta_{ij}+[\psi_\parallel(r)-\psi_\perp(r)]\frac{r_i r_j}{r^2}\ .
\ee
Thus, the velocity correlation/variogram is anisotropic on all scales
 (also see Fig. 7 and the related description by \citet{zhengyi13}).
Considering this anisotropy in the velocity correlation/variogram,
 in principle, we could Kriging interpolate one velocity component from all the three velocity components of other points.
This is an issue for future investigation.

Toward small scales, the velocity variogram is inhomogeneous for significant environment dependence.
For example, the velocity difference is small along a filamentary structure and large in the perpendicular directions.
This anisotropic variogram depends on the detailed environment.
Thus, for better reconstruction of the small-scale velocity field, we expect better performance with a position dependent anisotropic variogram.

It is interesting to test the performance of Kriging with an inhomogeneous and anisotropic variogram.
However, the main focus of this paper is the performance test of the most straightforward application of Kriging
 to measure the volume-weighted velocity on scale $k\sim 0.1\hmpc$ as the first step.
We will discuss the possible extensions by adopting an inhomogeneous and anisotropic variogram in Sec. \ref{sec:conclusion} and leave them for future work.


\section{Simulation Specification}
\label{sec:simulation}

\subsection{Simulation}

The simulation is run by a particle-particle-particle-mesh code
 adopting standard flat Lambda cold dark matter ($\Lambda$CDM) 
 with $\Omega_m=0.268$, $\Omega_\Lambda=0.732$, $\sigma_8=0.85$, $n_s=1$, and $h=0.71$ (see Ref. \cite{jingyp07}).
There are $1024^3$ dark matter (DM) particles inside a box of size $1200\mpch$.
We randomly sample the DM particles in simulation with fractions of $f=100\%$, $10\%$, $1\%$, $0.1\%$ and $0.01\%$
 to see the dependence of the velocity statistics upon the sample density.
The sample density $n_P$ scales with $f$ as
\be
n_P=\frac{1024^3}{1200^3}f\ (\mpch)^{-3}\ .
\ee
For each case, we interpolate velocities on a $256^3$ uniform grid by the Kriging method from the nearest $n_k=200$ observed points.

In observations, the volume-weighted velocity field is interpolated from halo/galaxy velocities.
In principle, we should quantify the performance of Kriging on the halo field in simulation.
However, due to the unknown halo velocity bias, we are lacking the correct volume-weighted halo velocity field even in simulation.
While the number density of the full DM sample in this simulation is sufficiently large
 and the velocity field constructed from the full sample could be treated as the right answer \cite{zhengyi15a}.
Thus, we choose to quantify the performance of Kriging on randomly selected DM particles.
Although the randomly selected DM particles do not exactly share the same distribution as real haloes,
 this process keeps the clustering property of the matter in the Universe.
 
Specifically, $f=0.1\%$ and $1\%$ correspond to number densities of $n_P\sim 6\times10^{-4} (\mpch)^{-3}$ and $6\times10^{-3} (\mpch)^{-3}$.
A future wide spectroscopic survey like Euclid\footnote{http://sci.esa.int/euclid/} and DESI\footnote{http://desi.lbl.gov/} could measure velocity data with this number density.
Thus, the performance of Kriging for these two cases is of particular interest.


\subsection{Statistics}
\label{sec:statistics}

Any vector field can be decomposed into an irrotational (gradient) part and a rotational (curl) part.
Analogous to the electric and magnetic fields, we denote the first one with a subscript``E" and the later one with a subscript ``B"
\footnote{In the literature, this decomposition is also denoted as $\bmv=\bmv_\parallel+\bmv_\perp$.
In Fourier space, the irrotational part $\bmv_\parallel$ is the component parallel to the vector $\bmk$,
 while the rotational part $\bmv_\perp$ is the one perpendicular to $\bmk$.}.
Decomposing the cosmic velocity field into the irrotational E-mode component and rotational B-mode component
 is widely used in cosmology studies.
 
During linear evolution, the cosmic velocity field only has an E-mode component.
Thus, at the early time or on the large scales, the velocity field is relatively easy to model.
It could be completely described by its divergence, which is related to the matter density field \cite{Peebles80}.
Late-time nonlinear evolution induces shell crossing in the cosmic flow, producing the B-mode component on small scales \cite{Pueblas09}.
Both components are related to the matter growth rate of the universe, thus carrying cosmological information.

Stage IV dark energy projects, such as MS-DESI (BigBOSS), Euclid, SKA, and WFIRST,
 can measure the volume-weighted velocity power spectrum through RSD with $\mathcal{O}(1\%)$ statistical precision (e.g., Ref. \cite{BigBOSS11}).
RSD is induced by a peculiar velocity.
Thus, for RSD modelling, it is of crucial importance to understand the peculiar velocity field.
An appropriate velocity decomposition has the potential to simplify and improve the RSD modelling \cite{zhangpj13}.
As the first step, the velocity is decomposed into an E-mode and B-mode component.
The E-mode component is responsible for the large-scale Kaiser effect \cite{Kaiser87}.

In this study, we focus on the E-mode component, which contains most of the cosmological information.
First, the volume-weighted velocity field is Kriging interpolated on regular grids.
Then, the interpolated velocity field is decomposed into an E-mode and B-mode component,
\be
\bmvstar=\bmve+\bmvb\ ,
\ee
in which the E-mode component is curl free ($\nabla\times\bmve=\bm{0}$)
and the B-mode component is divergence free ($\nabla\cdot\bmvb=0$).
This decomposition could be done using a fast Fourier Transform,
\ba
\bmve=\frac{(\bmvstar\cdot\bmk)\bmk}{k^2}\ ,\\
\bmvb=\bmvstar-\bmve\ .
\ea
We measure the E-mode velocity power spectrum
\be
P_{\bmv\bmv}(k)=\langle\bmve\bmve^*\rangle\ ,
\ee
and usually we present the result in the form of $\Delta^2_{\bmv\bmv}(k)=k^3P_{\bmv\bmv}(k)/2\pi^2$.

We also obtain the same statistics for the velocity field obtained by the NP method.
According to the convergence test by \citet{zhengyi15a}, we consider the result from the NP method with $f=100\%$ as a reference.


\section{Performance of Kriging}
\label{sec:results}

The performance of Kriging depends on the variogram prior $\gamma(r)$,
 the number $n_k$ of the observed points used to interpolate, 
 and the density of the observed sample described by the sampling fraction $f$.
Notice that the sampling fraction $f$ (or equivalently, the sample density $n_P$) is not an input parameter in the Kriging method.
When dealing with all dark matter particles in our simulation, $f=100\%$.
For massive and small haloes in observation, the number density corresponds to $f\sim 0.1\%$ and $\sim 1\%$, respectively.
It is a fixed number when the observed sample is determined.
However, it has a great impact on the performance of Kriging.
Thus, we present the dependence on $f$ thorough the whole performance test.

First, we quantify the performance of Kriging in the case that a theoretically predicted variogram is taken as the prior.
Then, the dependence on $n_k$ is shown as a convergence test.
Finally, the sensitivity on the variogram prior is tested by adopting inconsistent variograms.

\begin{figure}
\epsfxsize=8cm
\epsffile{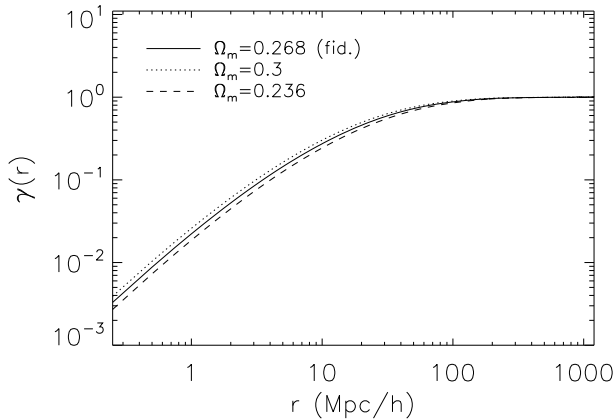}
\caption{Theoretically predicted variogram for the cosmological parameter used in simulation (solid line)
 and the variogram from a $\Omega_m=0.3$ and $\Omega_m=0.236$ flat universe (dotted line and dashed line, respectively).
\label{fig:varis}}
\end{figure}

\begin{figure*}
\epsfxsize=8cm
\epsffile{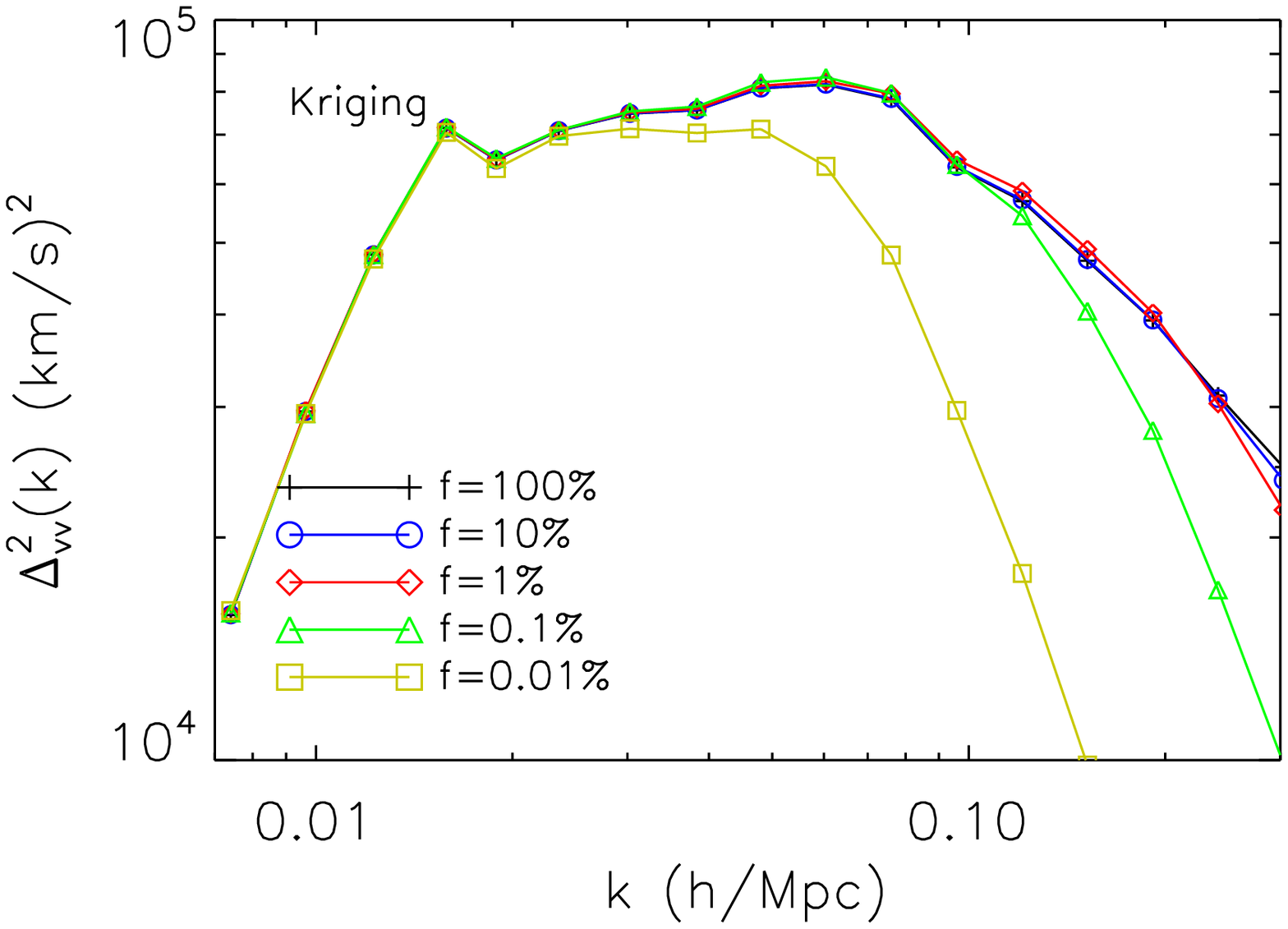}
\epsfxsize=8cm
\epsffile{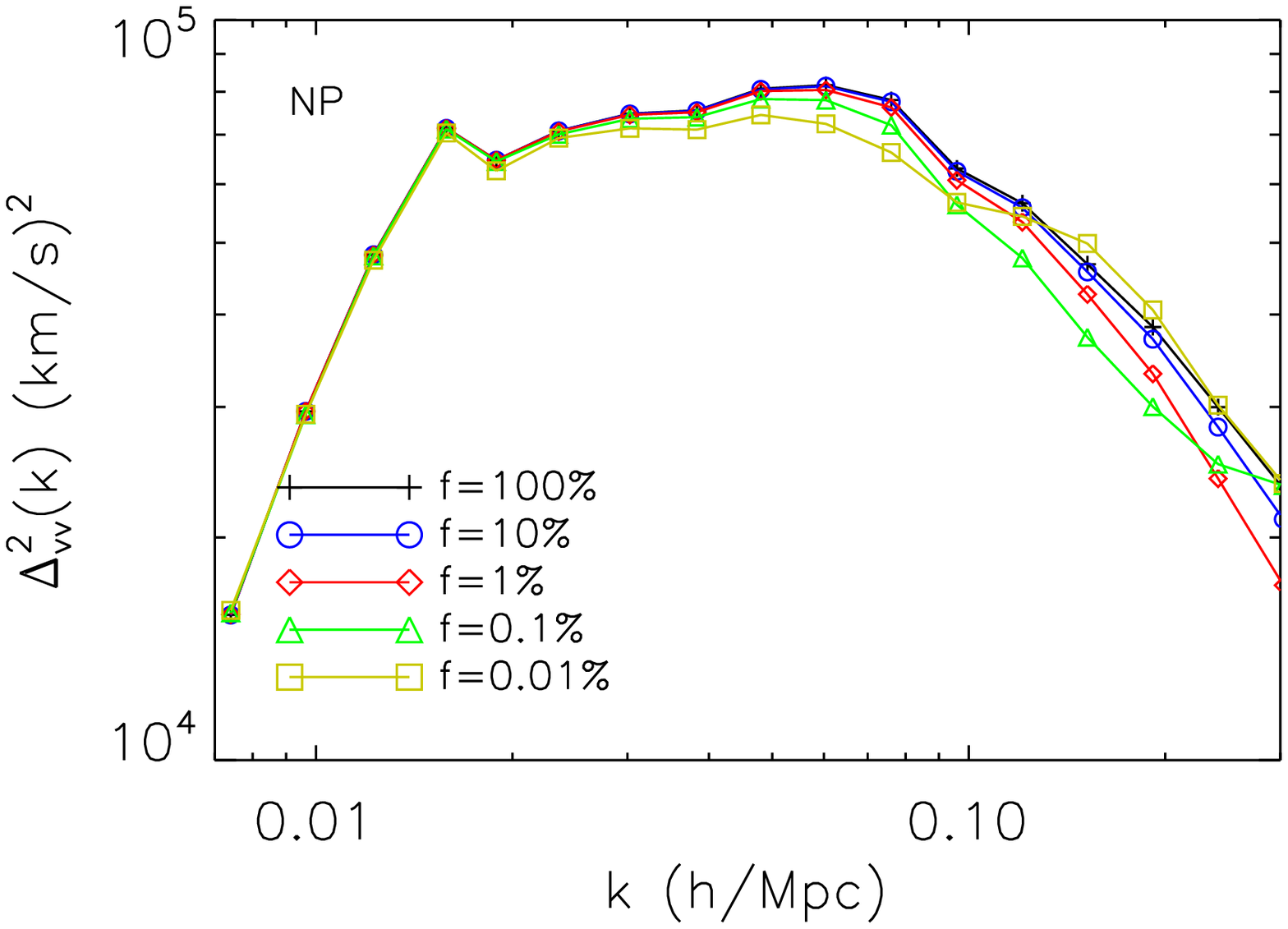}
\caption{E-mode velocity power spectrum from the Kriging method is presented in the left panel in which $n_k=200$ is adopted.
The variogram prior in the Kriging method is predicted by the \texttt{CLASS} code.
The one from the NP method is presented in the right panel.
\label{fig:vekgyu}}
\end{figure*}

\begin{figure*}
\epsfxsize=8cm
\epsffile{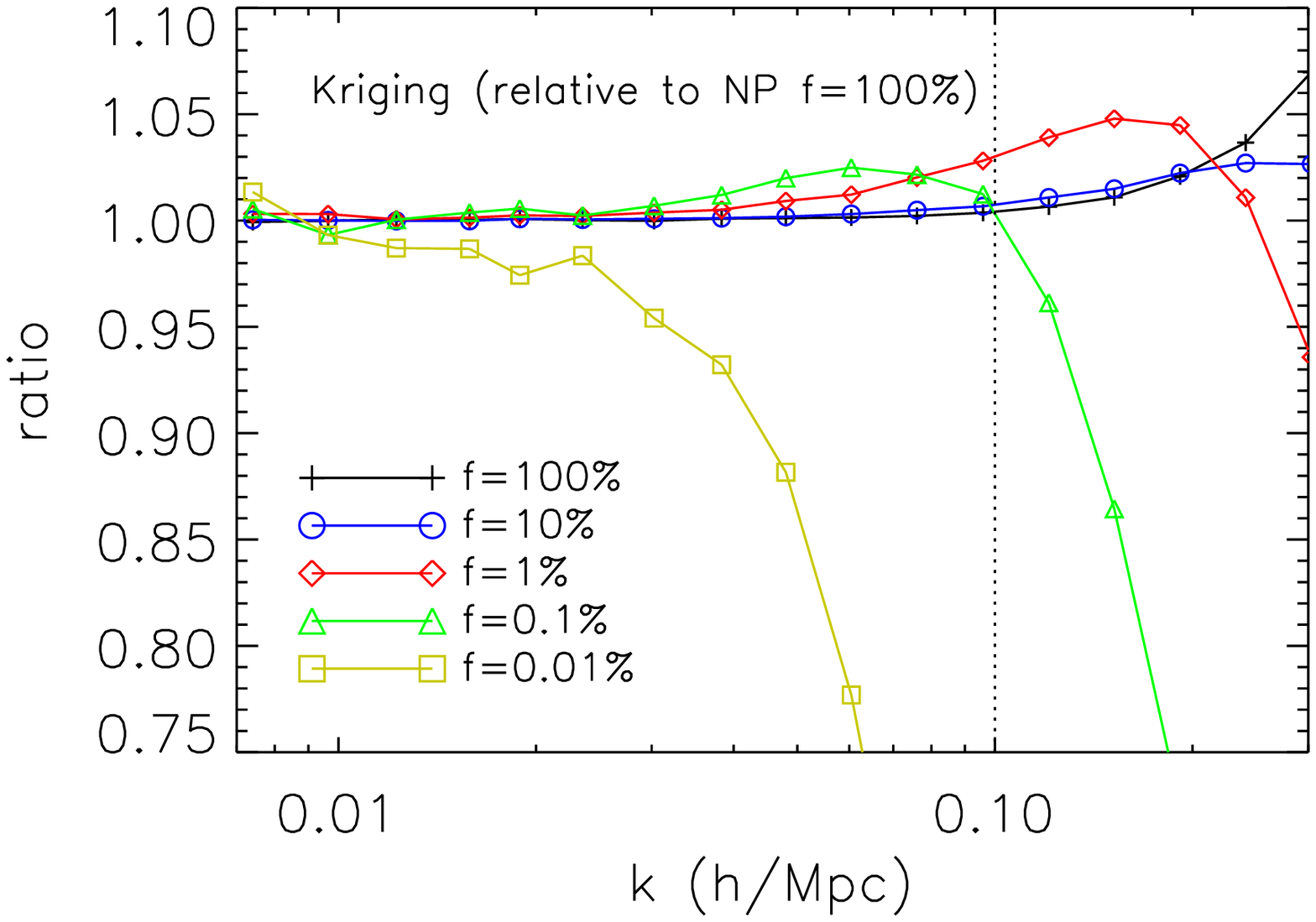}
\epsfxsize=8cm
\epsffile{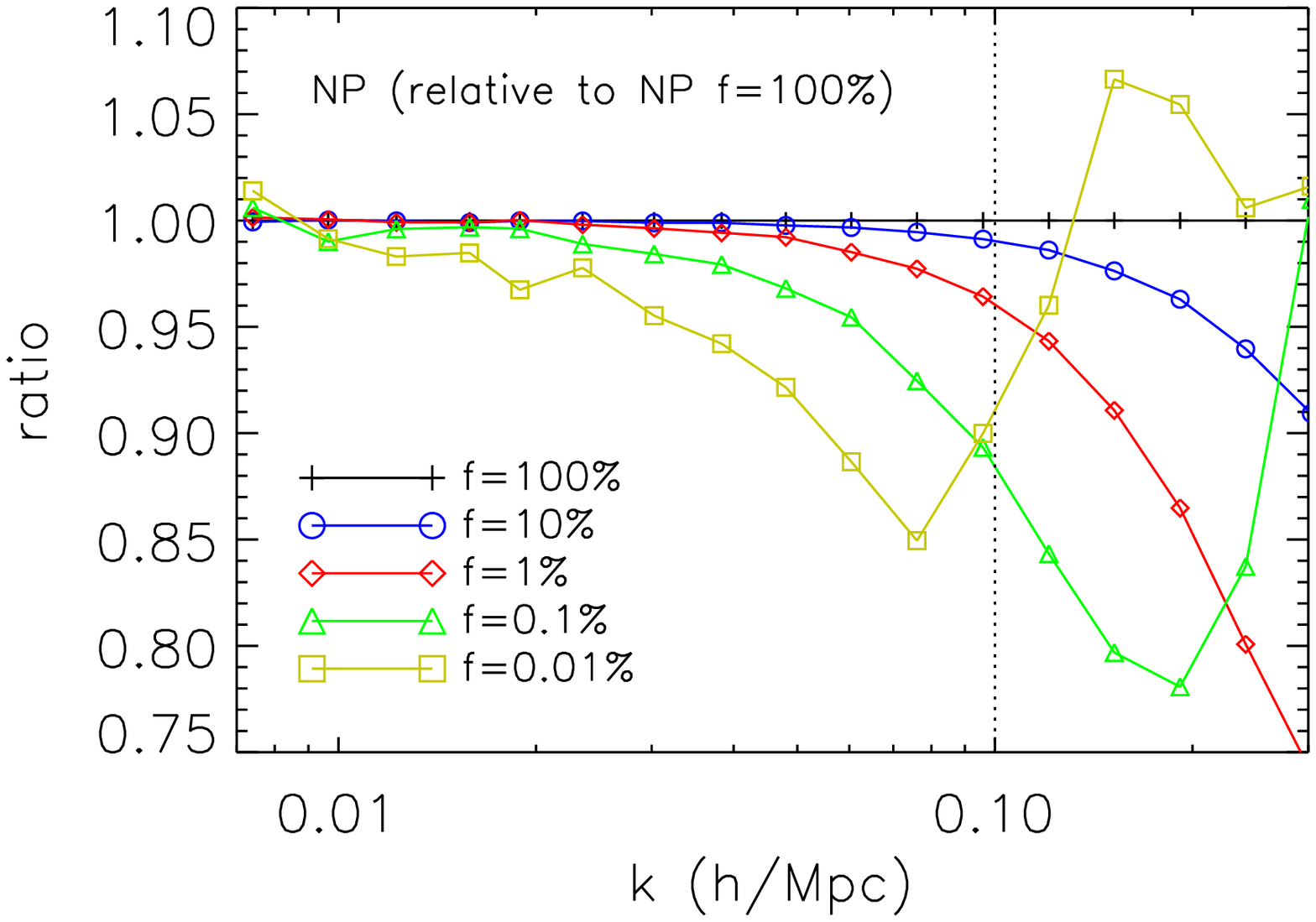}
\caption{Ratio of E-mode velocity power spectrum from the Kriging method with $n_k=200$ (left panel) and the NP method (right panel) to the reference one.
The vertical dotted lines indicate the most concerned scale $k=0.1\hmpc$.
Notice that for $f=100\%$ Kriging and NP agree with each other well within $1\%$ at $k\le 0.1\hmpc$ (left panel).
This confirms the previous conclusion that the NP method is robust at $k\le0.1\hmpc$ when $n_P\gtrsim 0.6(\mpch)^{-3}$.
\label{fig:vekgyupsratio}}
\end{figure*}

\subsection{Theoretically predicted variogram}
\label{sec:classvariogram}

We use the \texttt{CLASS} code \footnote{http://class-code.net} to produce the velocity divergence power spectrum
 $\Delta^2_{\theta\theta}(k)$. 
The correlation function is the Fourier pair of the power spectrum,
\be
\xi_{\bmv\bmv}(r)=\int\dif \ln k \frac{\Delta^2_{\theta\theta}(k)}{k^2}\frac{\sin(kr)}{kr}\ .
\ee
Since the absolute amplitude of the variogram does not influence Kriging interpolation (see Appendix. \ref{app:normalization}),
 we set the variogram as
\be
\gamma(r)=1-\xi_{\bmv\bmv}(r)/\xi_{\bmv\bmv}(0)\ .
\ee
This relation holds only if $\langle \bmv(\bm{x})\rangle=0$, true for the velocity.
The theoretically predicted variogram is presented in Fig. \ref{fig:varis} as a solid line.
It has a slope of zero beyond the velocity correlation length $r\sim 150\mpch$.
The most inner part of the variogram has slope of $2$ by the definition of the correlation function,
\be
\begin{split}
\xi_{\bmv\bmv}(r)&=\int\dif\ln k P_{\bmv\bmv}(k) \frac{\sin(kr)}{kr}\\
&=\int\dif\ln k P_{\bmv\bmv}(k) (1-\frac{1}{3!}(kr)^2+\mathcal{O}((kr)^4))\\
&=\xi_{\bmv\bmv}(0)-\frac{r^2}{6}\int\dif\ln kP_{\bmv\bmv}(k)k^2+\cdots\ ,
\end{split}
\ee
in which Taylor expansion is adopted in the second line.

The right panel of Fig. \ref{fig:vekgyu} is just the same result as Fig. 1 by \citet{zhengyi15a}.
For the NP method, it causes a $\sim 12\%$ underestimation of the velocity power spectrum at $k = 0.1\hmpc$ for the $f=0.1\%$ case.
This systematic underestimation increases with decreasing $f$ and increasing $k$.
However, also reported is an anomalous behavior, which is not expected from simple modelling of the sampling artifacts for the NP method.
For the $f=0.01\%$ case, the decreasing power turns up at small scales $(k>0.08 \hmpc)$.
Two possibilities causing this anomaly are discussed in the appendix of \citet{zhengyi15a}.
We refer the readers to \citet{zhangpj15} and \citet{zhengyi15a} for the detailed sampling artifacts study and modelling for the NP method.

Adopting the theoretically predicted variogram, the E-mode power spectrum obtained from the Kriging method with $n_k=200$
 is presented in the left panel of Fig. \ref{fig:vekgyu}.
We find a well-reconstructed power spectrum for $f=100\%$, $10\%$, and $1\%$.
Obvious deviation only appears for very low sampling fractions $f=0.1\%$ and $0.01\%$.
For $f=0.1\%$, the power spectrum is suppressed at $k>0.1\hmpc$,
 while for $f=0.01\%$, the suppression begins at a much larger scale ($k>0.02\hmpc$).
The success for $f=100\%$ to $1\%$ implies that this theoretically predicted variogram is reasonably close to the true one,
 although it neglects the nonlinear evolution of the velocity.
The sudden failure for $f=0.1\%$ and $0.01\%$ shows that the sampling fraction has a great impact on the performance of Kriging.

The mean particle separation $L_P$ scales with $f$ as 
\be
L_P=\frac{1200}{1024} f^{-1/3}\ \mpch\ .
\ee
For $f=0.1\%$ and $0.01\%$, the mean particle separation is $11.72\mpch$ and $25.25\mpch$, respectively.
The mean distance of one grid point to its nearest particle measured in simulation is $7.77\mpch$ and $15.55\mpch$, respectively.
From Fig. \ref{fig:varis}, we can see that for these separations of the $f=0.1\%$ case the variogram has values of $\gamma_{ij}\gtrsim 0.35$ and $\gamma_{i*}\gtrsim 0.2$.
This means that the velocity difference among nearby particles and between the grid point and nearby particles is large.
Thus, in these situations, all the nearby particles obtain comparable weightings from the Kriging system,
 leading to a large smoothing effect.

We define a characteristic length $L_v$ as
\be
L_v=\left( \frac{\langle(\nabla\cdot\bmv)^2\rangle}{\langle\bmv^2\rangle} \right)^{-\frac{1}{2}}\ .
\ee
This characteristic length indicates on which scale the velocity varies significantly.
If the mean particle separation $L_P$ is larger than this characteristic scale $L_v$, the velocity variation is not well sampled.
$L_v$ measured from the velocity field obtained by the NP method with $f=100\%$ is $L_v\sim 6.64\mpch$.
For $f=100\%$, $10\%$ and $1\%$, $L_P=1.17$, $2.52$ and $5.44\mpch$, smaller than $L_v$.
For these cases, the observed sample is sufficiently dense, and Kriging has good performance,
For $f=0.1\%$ and $0.01\%$, $L_P>L_v$.
In these situations, the velocity field is not well sampled, and therefore a smoothing effect dominates.

We present the ratio of the E-mode velocity power spectrum from the Kriging method and NP method to the reference one
 in the left and right panels of Fig. \ref{fig:vekgyupsratio}.
According to the modelling of the sampling artifacts for the NP method by \citet{zhangpj15},
 the reference case only suffers from negligible sampling artifacts at scales $k<0.2\hmpc$ (see Fig.3 by \citet{zhangpj15}).
The simulation results in \citet{zhengyi15a} show that the sampling artifacts of $f=10\%$ are only $1\%$ at $k=0.1\hmpc$.
This also implies negligible sampling artifacts for $f=100\%$.
Thus, the reference case is well approximated as the correct answer at the scales we consider.
The right panel of Fig. \ref{fig:vekgyupsratio} is the same result as Fig. 2 by \citet{zhengyi15a}.

We find that at scales larger than the smoothing dominating region power from the Kriging method is slightly larger than the reference one.
The nonlinear velocity power spectrum is predicted and found to be lower than the linear prediction at small scales (e.g., Ref. \cite{Bernardeau02b} for a perturbative prediction and Refs. \cite{Pueblas09,Hahn14,Inman15} for the simulation result).
Thus, we expect the true variogram to be lower than the linear prediction at small scales.  
The deviation of the theoretically predicted variogram from the true one may cause this overestimation in the power spectrum of the Kriging interpolated velocity field.

Compared to the NP method, the most straightforward Kriging does not show obvious advantages.
For $f \gtrsim 1\%$, the NP method underestimates the small-scale power more with decreasing $f$.
This is an expected behavior since for lower $f$ the NP method assigns the same velocity to more grids,
 leading to a larger smoothing effect,
while for the Kriging method, the measured power is overestimated at intermediate scales and suppressed at small scales.
Considering that the overestimation depends on the variogram prior and we only use a linear prediction,
 a detailed comparison is less meaningful.
In fact, in the following analysis (Sec. \ref{sec:cosmo}), we find that the power amplitude depends on the cosmological parameters used to predict linear variogram prior.
This implies that the inaccuracy of the variogram prior could affect the performance of Kriging at the percent level.

For $f=0.1\%$ and $0.01\%$, a heavy smoothing effect dominates the performance of Kriging,
while for the NP method, these two situations also produce anomalous behavior, which is difficult to model.
We emphasize that the sudden failure in Kriging and anomalous behavior in the NP method is physical.
For such low sampling faction, the velocity field is not well sampled, and we do not expect good performance for any velocity assignment method.

Notice that for $f \gtrsim 1\%$ the sampling artifacts of the NP method are relatively straightforward to model and to some extent it could be corrected.
Although the sampling artifacts are difficult to model for the more complicated Kriging method,
 we expect improvement by applying a more delicate version of Kriging.
We discuss the potential improvement that could be made in Sec. \ref{sec:conclusion} and leave this for a  futrue investigation.

\begin{figure*}
\epsfxsize=16cm
\epsffile{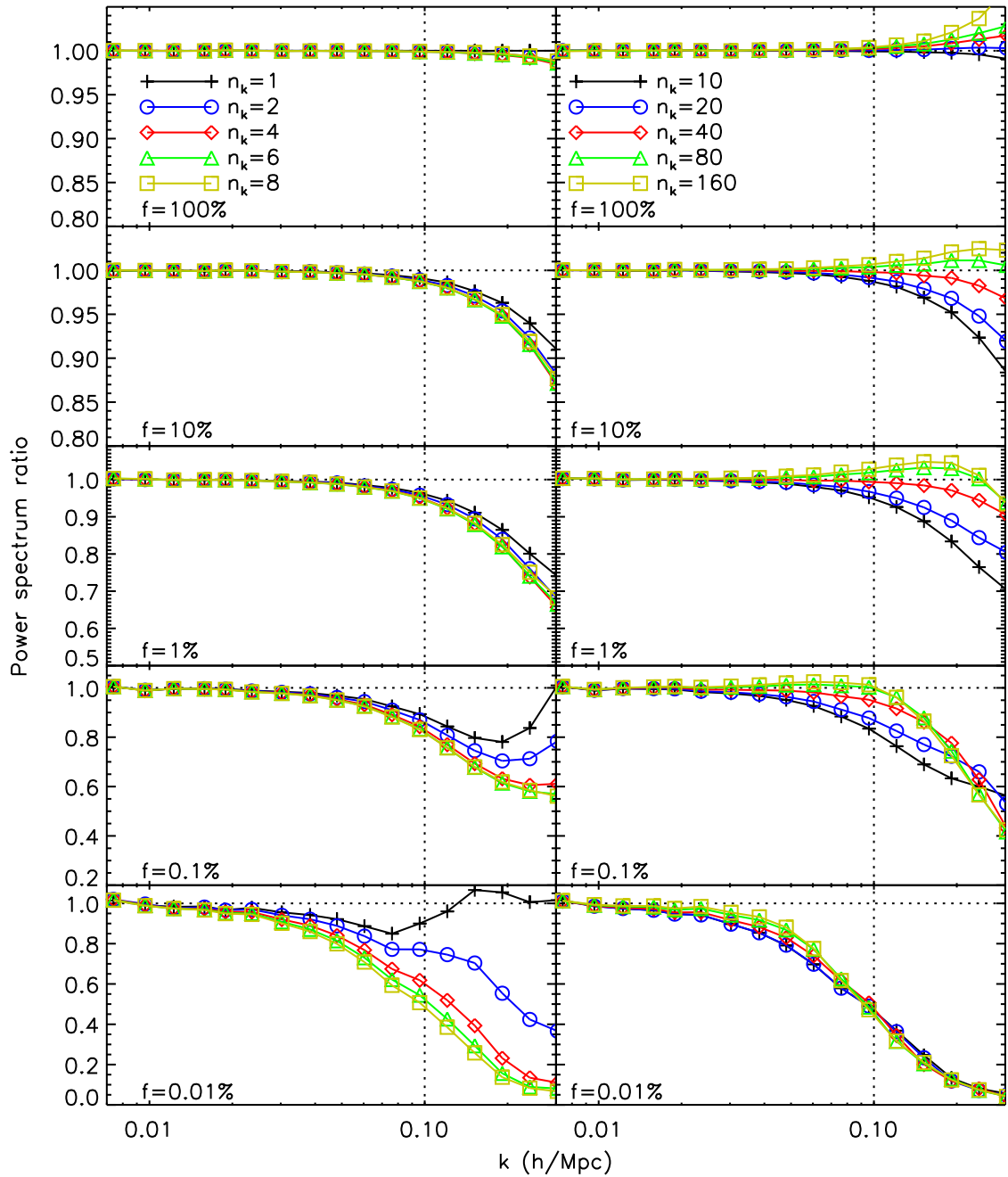}
\caption{
Ratio of the E-mode velocity power spectrum from the Kriging method to the reference one.
The number of particles to Kriging interpolate increases from $n_k=1$ to $8$ (left panel) and $n_k=10$ to $160$ (right panel).
From top to bottom, the sampling fraction $f$ decreases from $100\%$ to $0.01\%$.
The vertical dotted lines indicate the most concerned scale $k=0.1\hmpc$.
Note that the y-axis scaling is different to better show the results.
\label{fig:nk}}
\end{figure*}

\begin{figure*}
\epsfxsize=8cm
\epsffile{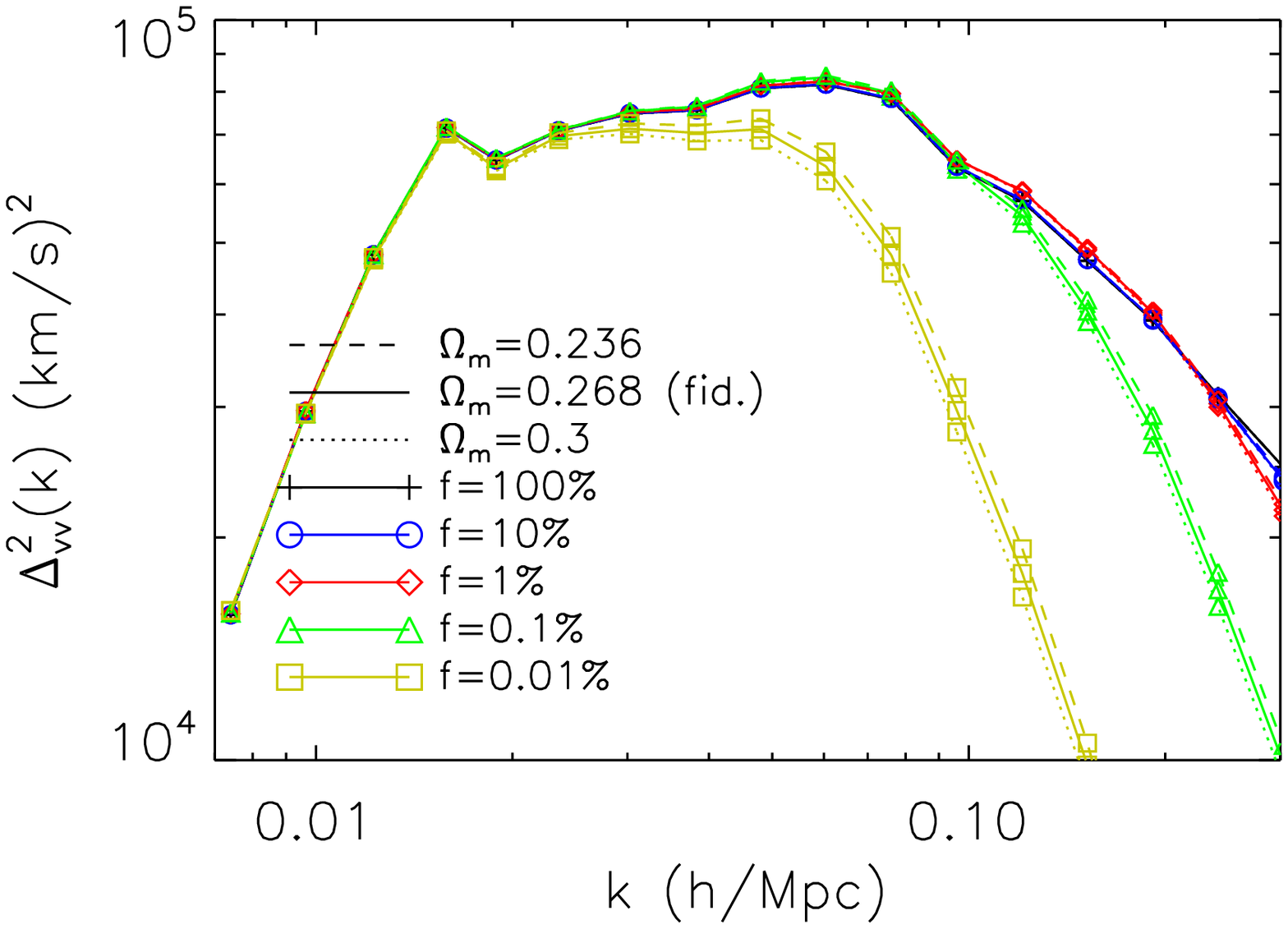}
\epsfxsize=8cm
\epsffile{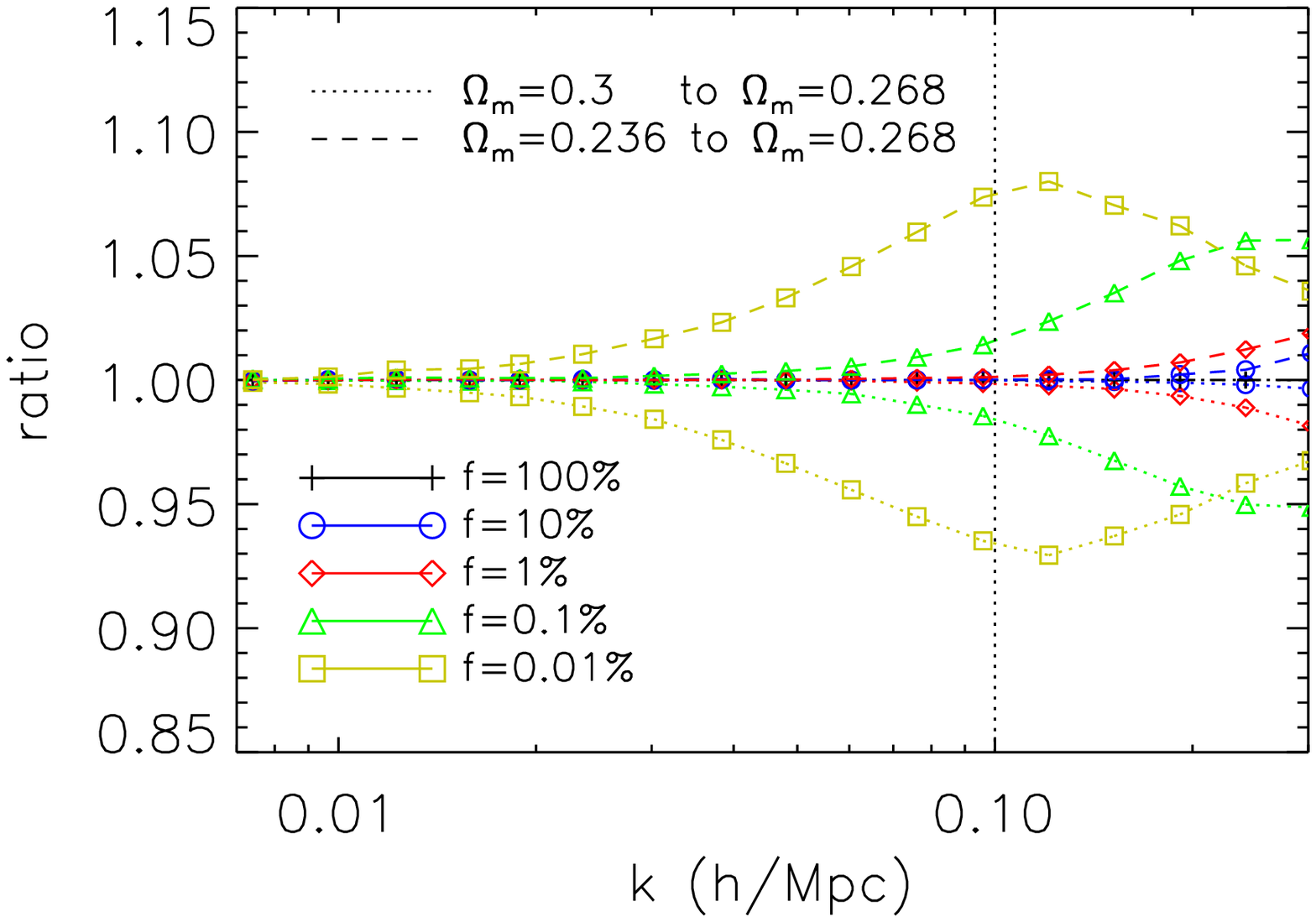}
\caption{E-mode velocity power spectrum from the Kriging method using a consistent prior (solid line), and
 inconsistent ones (dotted and dashed line) are presented in the left panel.
The ratio to the fiducial one is presented in the right panel.
The vertical dotted line indicates the most concerned scale $k=0.1\hmpc$.
\label{fig:vekgcosmo}}
\end{figure*}

\subsection{Dependence on $n_k$}
\label{sec:convergence}

In Fig. \ref{fig:varis}, we can see that the variogram approaches to unity at $\sim 150\mpch$.
This implies that particles with distance larger than $\sim 150\mpch$ do not contribute to the prediction of the velocity on a given grid point.
Thus, in principle, for each grid point, we only need to include all the particles with a distance less than $\sim 150\mpch$.
However, for high sampling fraction cases, this implementation is still computationally expensive.
Closer particles generally have a larger contribution to the estimation of the velocity on the grid point.
Thus, we naturally choose to use the nearest $n_k$ particles from the grid point to Kriging interpolate.
Which $n_k$ is suitable relies on many factors.

One can image an extreme case with infinite particles in simulation.
For any grid point, there always exists a particle passing it.
Since Kriging is accurate in this case, $n_k=1$ (i.e., reduced to the NP method) is enough.
For general cases, whether the outer-part particles are important depends on the steepness of variogram.
A very steep inner part of the variogram means that the velocity on the grid is very likely to be similar as close particles.
Thus, the weightings for outer-part particles are very small.
In this case, excluding a part of the outer particles will not influence the interpolation result.

The performance test with increasing $n_k$ will greatly help us understand the Kriging method.
Figure \ref{fig:nk} presents the dependence on the particle number $n_k$ used to Kriging interpolate.
In the left plots, $n_k$ increases from 1 to 8, and in the right plots, it increases from 10 to 160.
The ratio of the interpolated velocity power to the reference one is presented to better show the dependence.

For $f=100\%$, all the lines overlap with each other when $k<0.1\hmpc$.
This implies that for such a high density of particles the velocity on the grid could be determined well by only a few nearby particles.
This result is consistent with the conclusion by \citet{zhengyi15a} that for $f=100\%$ the NP method is accurate.
For $f=10\%$ and $1\%$, underestimation at small scales could be seen when only a few particles are used in interpolation.
However, further increasing $n_k$ increases the small-scale power toward (even over) the reference one.
This behavior could be understood as follows.
Inside a low-density region, several grid points share the same nearest particle, obtaining the same velocity from the NP method.
Using only a few particles to Kriging interpolate also suffers from a similar sampling artifact.
This sampling artifact assigns a similar velocity to several nearby grid points, therefore suppressing the small-scale power.
Including more particles to interpolate gathers more information.
This helps to recover the difference between the nearby grid points, leading to the recovery of small-scale power.

For $f=1\%$, $0.1\%$, and $0.01\%$, increasing $n_k$ from 1 to 8 gradually suppresses the power more.
This reflects the smoothing effect of the Kriging method.
However, further increasing $n_k$ from 10 to 160 recovers the small-scale power for $f=1\%$ and
 increases the power at $0.02<k<0.2\hmpc$ for $f=0.1\%$ and $0.01<k<0.1\hmpc$ for $f=0.01\%$.
This also reflects the help by including more particles to interpolate.

For $f=10\%$ and $1\%$, increasing $n_k$ to 160 almost fully recovers the power at all scales we consider.
But for $f=0.1\%$, the power at scales $k>0.1\hmpc$ is still underestimated even $n_k=200$.
Further increasing $n_k$ will not help since new added outer-part particles are far from the grid point.
Thus, this small-scale power is lost in the Kriging method.
For $f=0.01\%$, the situation is even worse.
The power is underestimated at $k>0.015\hmpc$ when all particles are used.

For $f<1\%$ cases, increasing $n_k$ from 80 to 160 does not change the power significantly.
Thus, the performance of Kriging converges at $n_k\sim 100$.

Notice that the observed velocity sample has spatial clustering.
Particles with coherent motion inside a small dense region do not provide independent information to Kriging system.
Although this will not influence the result, it reduces the performance of Kriging in the aspect of a waste of computational resources.
For low sampling fraction cases like the halo population in our Universe,
 the mean separation is large, and we do not expect a large degradation.
For a high sampling fraction, a preselected sample with fairer spatial distribution could improve the efficiency of Kriging interpolation.

For the higher sampling fraction case, smaller $n_k$  is needed to reconstruct the power spectrum well.
Thus, we could expect that for denser regions smaller $n_k$ is needed.
Determining $n_k$ for each grid point from its density could also improve the computational efficiency.

\subsection{Sensitivity on the variogram prior}
\label{sec:cosmo}

For performance test on simulation, we know the true cosmology and could obtain a consistent theoretical variogram.
When applying Kriging to observations, we do not know the cosmological parameters.
Thus, we quantify the sensitivity on the variogram prior by adopting two inconsistent variograms in Kriging.
These two variograms are produced by modifying $\Omega_m$ from 0.268 to 0.3 and 0.236 while keeping the universe flat,
 since the structure formation is very sensitive to $\Omega_m$.

Two inconsistent variograms we adopted are presented in Fig. \ref{fig:varis}.
For larger (smaller) $\Omega_m$, the (scaled) variogram is slightly higher (lower) at the scales below the velocity correlation length.
The E-mode velocity power spectrum obtained by the Kriging method with these inconsistent variograms
 is presented in the left panel of Fig. \ref{fig:vekgcosmo}, and the ratio to the fiducial one is presented in the right panel.
We find that the deviation amplitude is larger with decreasing $f$.
An inconsistent variogram with larger (smaller) $\Omega_m$ produces an underestimated (overestimated) power spectrum.
The underestimation (overestimation) reaches $7-8\%$ for $f=0.01\%$ at $k\sim 0.1\hmpc$
 and $5\%$ for $f=0.1\%$ at $k\sim 0.3\hmpc$.

Notice that the power at the largest scale is insensitive to the variogram prior.
It implies that Kriging keeps the consistency between the linear evolution of the density and velocity field on large scales.
For decreasing $f$, the deviation appears at larger scales with a larger amplitude.
Thus, for lower sampling fractions, Kriging is more sensitive to the variogram prior.

The measured velocity power spectrum depends on the variogram prior, which itself depends on the cosmological parameters.
This implies that the Kriging interpolated velocity field from a sparse halo population could not be used to constrain cosmological parameters.
Thus, so far, we constrain the application of the Kriging method in the theoretical understanding of the peculiar velocity in simulation.

With good understanding of the peculiar velocity field and a good enough velocity variogram/correlation function model, 
one could try to fit the model parameter from the Kriging interpolated velocity field iteratively.
This process could produce a volume-weighted velocity field with a self-consistent variogram,
enabling the application for cosmology.
We tried this iterative method by adopting coarse variogram models.
We found the process is unstable since the resulting power depends significantly on the variogram model, especially on small scales, which is hard to accurately model.
This hampers the application of Kriging in cosmology, unless we could obtain a sufficiently good variogram model.


\section{Conclusion and Discussion}
\label{sec:conclusion}

For the first time, we applied the Kriging method to obtain the volume-weighted velocity field from N-body simulations.
We tested the performance of Kriging on the variogram prior $\gamma(r)$, the number $n_k$ of the nearby particles to interpolate, 
 and the density of the observed sample described by the sampling fraction $f$.
 
First, we theoretically predicted a linear variogram consistent with the simulation.
Adopting this theoretically predicted variogram, Kriging reconstructed the E-mode velocity power spectrum well for $f\ge 1\%$ at all the scales we consider.
However, for $f=0.1\%$ and $0.01\%$, a smoothing effect dominated small-scale power suppression.
This is a natural behavior when an observed sample is so sparse that the mean particle separation $L_P$ is larger than the characteristic scale $L_v$.
Second, we tested the $n_k$ dependence by increasing $n_k$ from 1 to 160.
We found that increasing $n_k$ from 1 to 8 generally smoothed the small-scale power first,
 while further increasing $n_k$ from 10 to 160 recovered the small-scale power to some extent.
For $f=10\%$ and $1\%$, the suppressed power could be recovered at all the scales we considered,
 while for lower $f$, the power was eventually lost even when more particles were included.
The Kriging system converged for $n_k\sim 100$.
Finally, we adopted inconsistent variograms to check the sensitivity on the variogram prior.
An inconsistent variogram biased the velocity power spectrum at small scales.
The deviation was larger for lower $f$ and began to appear at larger scales.
Adopting the variogram predicted by a $\Omega_m=0.3$ ($0.236$) flat universe, the underestimation (overestimation) of power reached $5\%$ at $k\sim 0.1\hmpc$ for $f=0.1\%$.

For $f>1\%$ and $f=0.1\%$ at $k<0.1\hmpc$, both the most straightforward Kriging method and NP method generally performed at comparable levels.
For lower $f$ cases, both methods could not reconstruct the velocity field well.
The Kriging method suddenly failed for these lower $f$ cases, and the power suffered a heavy smoothing effect,
while the NP method suffered from an unexpected behavior from simple sampling artifacts modelling, and was thus difficult to correct.

There are several potential improvements to be made for using the Kriging method
 to obtain the volume-weighted cosmic velocity field in the future.
First, as stated in Sec. \ref{sec:convergence}, a better strategy for selecting the observed sample
 and determining the number of points used to interpolate could improve the computational efficiency.
Second, as discussed in Sec. \ref{sec:velocitykriging}, due to the fact that velocity field is a vector field,
the correlation function/variogram is anisotropic on all scales.
Adopting the linear predicted but anisotropic variogram has the potential to improve the volume-weighted velocity reconstruction.
In fact, after some calculations, we did find improvement by considering the anisotropy in the variogram.
We expect this improvement to the Kriging method to be the most straightforward and potential one.
Third, to better reconstruct the velocity on small scales ($k>0.1\hmpc$), taking the position-dependent anisotropic variogram into account is promising.
To do this, we need understanding and modelling on the velocity for different cosmic structures (void, sheet, filament, and cluster).
Position-dependent Kriging interpolation is expected to suppress the smoothing effect for partially collapsed structures, leading to better small-scale reconstruction.
Fourth, taking the correlation between all the velocity components into account, 
 we could extend the Kriging method to interpolate each velocity component from all the velocity components.
In fact, under the approximation that the velocity is a gradient field,
 one can reconstruct one of the velocity components from another component equally as well as from the same component.
Thus, whether this extension could better reconstruct the velocity field is worth exploration.
Finally, we adopted the velocity variogram from the linear prediction, neglecting the nonlinear effect that is difficult to model.
More knowledge on the velocity variogram (or equivalently, the velocity correlation function) would help us to better reconstruct the cosmic velocity field.
Most important of all, good modelling of the velocity variogram, especially on small scales, would enable us to fit a consistent variogram from the interpolated field iteratively.
This would avoid the uncertainties induced by choosing the variogram prior.

Better understanding of the cosmic velocity field will improve the Kriging interpolation.
Conversely, to better understand the cosmic velocity field, one needs a good velocity assignment method with well-known or well-controlled sampling artifacts.
For example, protohalo statistics predicts the physical halo velocity bias ($b_v<1$)
 (e.g., Refs. \cite{Bardeen86,Desjacques08b,Desjacques10a}).
However, a profound assumption in peculiar velocity cosmology is $b_v=1$ at a sufficiently large scale.
Because of the severe sampling artifacts in measuring the volume-weighted velocity power spectrum for sparse populations,
 an accurate halo velocity bias measurement is hampered \cite{Biagetti14,zhengyi15b,Jennings15}.
After appropriate corrections of the sampling artifact in the NP method,
 \citet{zhengyi15b} verified $b_v=1$ at $k\lesssim 0.1\hmpc$ and $z=0-2$ for haloes of mass $\sim 10^{12}-10^{13}\Msunh$.
Nevertheless, the measured velocity bias is only accurate to a few percent due to the residual sampling artifact.
Better correction of sampling artifacts is needed to measure $b_v$ with $1\%$ accuracy.

In this paper, we only quantified the performance of Kriging through the E-mode velocity power spectrum.
The sampling artifact also affects other velocity statistics.
For example, \citet{zhangpj15} proved in the Appendix that the sampling artifact also affects the density-velocity correlation measured by the NP method.
A similar sampling artifact is also expected in that measured by the DT method \cite{Jennings15b}.
Although for the dense DM population that \citet{Jennings15b} dealt with the sampling artifact is negligible,
it must be appropriately corrected for a sparse halo population.
Thus, the performance of Kriging on other statistics (e.g., the divergence, velocity shear, and B-mode component on small scales) is also interesting to explore.
To study the B-mode velocity component, high-resolution analysis is essential since the B mode is significant on small scales and the correlation length is much shorter.
To what extent Kriging could help us to better understand the cosmic velocity field is worth further investigation.


\section*{Acknowledgments}

We thank Yi Zheng, Han Miao, and De-Chang Dai for useful discussions.
This work was supported by the National Science Foundation of China (Grants No. 11403071, No. 11025316, No. 11320101002, No. 11433001, and No. 11273018), 
National Basic Research Program of China (973 Programs No. 2013CB834900 and No. 2015CB857001), 
the Strategic Priority Research Program ``The Emergence of Cosmological Structures'' of the Chinese Academy of Sciences (Grant No. XDB09000000),
 and the key laboratory grant from the Office of Science and Technology, Shanghai Municipal Government (Grant No. 11DZ2260700).
 J.Z. is supported by the national Thousand Talents Program
for distinguished young scholars and the T.D. Lee Scholarship from the High Energy Physics Center of Peking University.
This work made use of the High Performance Computing Resource in the Core
Facility for Advanced Research Computing at Shanghai Astronomical Observatory.


\appendix
\numberwithin{equation}{section}

\section{Kriging interpolation}
\label{app:krigingproperty}

\subsection{Derivation}
\label{app:derivation}

We assume that we want to estimate the field at a given position $\bmx_*$ from $n_k$ observed data points at $\bmx_i$
 with value of $y_i=y(\bmx_i)$.
Kriging looks for the value of the field at this position as a weighted linear combination of the nearby values at known positions,
\be
\hat{y}(\bmx_*)=\sum_i W_i y_i\ .
\ee
The error in estimating $y(\bmx_*)$ is $\epsilon(\bmx_*)=\hat{y}(\bmx_*)-y(\bmx_*)$.
Since the estimator is modelled by a Gaussian process, to ensure that the model is unbiased, the unbiased constraint is observed,
\be
E(\epsilon(\bmx_*))=0\ ,
\ee
which leads to the unbiased condition $\bm{1^T}\cdot\bm{W}=1$.

To find an estimator with minimum variance, we need to minimize $E(\epsilon^2(\bmx_*))$:
\be
\begin{split}
E(\epsilon^2(\bmx_*))&=\bm{W}^T\cdot\bm{Cov}(\bmx_i\bmx_j)\cdot\bm{W}-\bm{Cov}^T(\bmx_i\bmx_*)\cdot\bm{W}\\
                                     &-\bm{W}^T\cdot\bm{Cov}(\bmx_i\bmx_*)+Var(\bmx_*)\ .
\end{split}
\ee
Under the assumption of isotropy, the elements in $\bm{Cov}(\bmx_i\bmx_j)$ are the correlation among the observed points $\xi(|\bmx_i-\bmx_j|)$,
 while the elements in $\bm{Cov}(\bmx_i\bmx_*)$ are the correlation among observed points and the interpolated point $\xi(|\bmx_i-\bmx_*|)$.

By the Lagrange multiplier method, solving this optimization problem results in the Kriging system
\be
\left[\begin{array}{c} \bm{W} \\ \mu \end{array}\right]
=\left[\begin{array}{cc} \bm{Cov}(\bmx_i\bmx_j) & \bm{1} \\ \bm{1^T} & 0 \end{array}\right]^{-1}
\left[\begin{array}{c} \bm{Cov}(\bmx_i\bmx_*) \\ 1 \end{array}\right]\ .
\ee
Adopting the relation between the variogram and correlation $\gamma(r)=\xi(0)-\xi(r)$, usually it is expressed in terms of the variogram:
\be
\left[\begin{array}{c} \bm{W} \\ \mu \end{array}\right]
=\left[\begin{array}{cc} \bm{\gamma}_{ij} & \bm{1} \\ \bm{1^T} & 0 \end{array}\right]^{-1}
\left[\begin{array}{c} \bm{\gamma}_{i*} \\ 1 \end{array}\right]\ .
\ee
Here the elements in $\bm{\gamma}_{ij}$ are the variogram among the observed points $\gamma(|\bmx_i-\bmx_j|)$,
 while the elements in $\bm{\gamma}_{i*}$ are the variogram among the observed points and the interpolated point $\gamma(|\bmx_i-\bmx_*|)$.

\subsection{Trivial cases}
\label{app:trivial}

If only the nearest particle is used in Kriging, i.e., $n_k=1$, we have
\be
\left[\begin{array}{c}w\\ \mu\end{array}\right]=
\left[\begin{array}{ccc} 0 & 1 \\  1 & 0  \end{array}\right]^{-1}
\left[\begin{array}{c} \gamma_{1*} \\ 1\end{array}\right]=
\left[\begin{array}{c} 1 \\ \gamma_{1*}\end{array}\right]\ .
\ee
In this case, it reduces to the NP method, i.e., setting the velocity on grid as the velocity of the nearest particle.

Considering $n_k=2$, we have
\be
\left[\begin{array}{c}w_1\\w_2\\ \mu\end{array}\right]=
\left[\begin{array}{ccc} 0 & \gamma_{12} & 1 \\ \gamma_{12} & 0 & 1 \\ 1 & 1 & 0  \end{array}\right]^{-1}
\left[\begin{array}{c} \gamma_{1*} \\ \gamma_{2*}\\ 1\end{array}\right]\ ,
\ee
\ba
w_1=\frac{1}{2}\frac{\gamma_{2*}-\gamma_{1*}+\gamma_{12}}{\gamma_{12}}\ ,\\
w_2=\frac{1}{2}\frac{\gamma_{1*}-\gamma_{2*}+\gamma_{12}}{\gamma_{12}}\ . 
\ea
In the case of the grid point approaching one of the particles (for example, particle 1),
 we have $\gamma_{1*}\rightarrow \gamma_{11}=0$, $\gamma_{2*}\rightarrow \gamma_{12}$, and $[w_1,w_2]\rightarrow [1, 0]$.
Thus, if very close to one of these particles, the grid point will just obtain the velocity of that particle.
In other words, the velocity assignment is accurate in this case, and the other particles are not important.
This is also true for any $n_k$.
It explains the result for the $f=100\%$ case in which the power spectrum is insensitive to $n_k$ (see Sec. \ref{sec:convergence}).

\subsection{Independence on the absolute amplitude of the variogram}
\label{app:normalization}

The velocity interpolated from the Kriging method only depends on the shape of the variogram (relative amplitude).
Considering multiplying the variogram by a factor of $\alpha$, the new weighting $\bm{\hat{W}}$ and Laplacian multiplier $\hat\mu$
 are solved from the new Kriging system
\be
\left[\begin{array}{cc} \alpha\bm{\gamma}_{ij} & \bm{1} \\ \bm{1^T} & 0 \end{array}\right]
\left[\begin{array}{c} \bm{\hat{W}} \\ \hat\mu \end{array}\right]
=\left[\begin{array}{c} \alpha\bm{\gamma}_{i*} \\ 1 \end{array}\right]\ .
\label{eqn:kriging2}
\ee
The first line reads
\be
\bm{\gamma_{ij}}\bm{\hat{W}}+\frac{1}{\alpha}\hat\mu\bm{1}=\bm{\gamma_{i*}}\ .
\ee
The second line is the unbiased condition
\be
\sum_i \hat{W}_i=1\ .
\ee
From the above two equations one could easily find that
the solution to this new Kriging system, Eq. (\ref{eqn:kriging2}), is the same as the solution to Eq. (\ref{eqn:kriging}); i.e., $\bm{\hat{W}}=\bm{W}$.
The only difference is the irrelevant Laplacian multiplier $\hat\mu=\alpha\mu$.

\subsection{Singular case}
\label{app:singular}

A general usage of Kriging interpolation is adopting a variogram fit from the same data used to interpolate.
For applications without a high-precision requirement, a power law fitting $\gamma=\alpha r^\beta$ is sufficient to obtain a reasonable good interpolated value.
However, for a strong linear case with $\gamma_{ij}=\frac{1}{2}|\bmv_i-\bmv_j|^2\propto r_{ij}^2$, this Kriging system is degenerated.  The proof is as follows:
\be
\begin{split}
&\det(\bm{A})
=\left|\begin{array}{cc} \bm{\gamma}_{ij} & \bm{1} \\ \bm{1^T} & 0 \end{array}\right|
=\left|\begin{array}{cc} |\bmr_i-\bmr_j|^2 & \bm{1} \\ \bm{1^T} & 0 \end{array}\right|\\
&=\left|\begin{array}{cc} \bmr_i^2+\bmr_j^2-2\bmr_i\cdot\bmr_j & \bm{1} \\ \bm{1^T} & 0 \end{array}\right|
=\left|\begin{array}{cc} -2\bmr_i\cdot\bmr_j & \bm{1} \\ \bm{1^T} & 0 \end{array}\right|\\
&=(-2)^{n_k}\left|\begin{array}{cc} \bmr_i\cdot\bmr_j & \bm{1} \\ -\frac{1}{2}\bm{1^T} & 0 \end{array}\right|\ .
\end{split}
\ee
For $n_k>3$, $\bmr_i=(x_i,y_i,z_i)$, we have
\be
\left|\begin{array}{c}\bmr_i\cdot\bmr_j\end{array}\right|
=\left| \begin{array}{cccc} x_1 & y_1 & z_1 & \bm{0^T} \\ 
                                         \vdots & \vdots & \vdots & \vdots  \\
                                         x_{n_k} & y_{n_k} & z_{n_k} & \bm{0^T}  \end{array}\right|
\left| \begin{array}{ccc} x_1 & \cdots & x_{n_k} \\  
                                       y_1 & \cdots & y_{n_k} \\  
                                       z_1 & \cdots & z_{n_k} \\ 
                                       \bm{0} & \cdots & \bm{0}  \end{array}\right|=0\ .
\ee
Thus, the right-bottom cofactor does not affect the determinant.
We just set it as 0.5,
\be
\begin{split}
&\det(\bm{A})
\propto\left|\begin{array}{cc} \bmr_i\cdot\bmr_j & \bm{1} \\ -\frac{1}{2}\bm{1^T} & \frac{1}{2} \end{array}\right|
=\left|\begin{array}{cc} 1+\bmr_i\cdot\bmr_j & \bm{1} \\ \bm{0^T} & \frac{1}{2} \end{array}\right| \\
&\propto\left|\begin{array}{cc} 1+\bmr_i\cdot\bmr_j & \bm{0} \\ \bm{0^T} & 1 \end{array}\right|
=\left|\begin{array}{c} 1+\bmr_i\cdot\bmr_j\end{array}\right| \\
&=\left| \begin{array}{ccccc} 1 & x_1 & y_1 & z_1 & \bm{0^T} \\ 
                                           \vdots & \vdots & \vdots & \vdots & \vdots  \\
                                           1 & x_{n_k} & y_{n_k} & z_{n_k} & \bm{0^T}  \end{array}\right|
\left| \begin{array}{ccc}     1   & \cdots & 1 \\
                                       x_1 & \cdots & x_{n_k} \\  
                                       y_1 & \cdots & y_{n_k} \\  
                                       z_1 & \cdots & z_{n_k} \\ 
                                       \bm{0} & \cdots & \bm{0}  \end{array}\right|\ ,
\end{split}
\ee
which is zero for $n_k>4$.
To conclude, in the three-dimensional case, if $\gamma(r)\propto r^2$, we could not use the Kriging method for $n_k>4$.

Readers may find that $\det(\bm{A})$ is just the Cayley-Menger determinant.
Mathematically, the hypervolume of a $n_k$ simplex is related to the Cayley-Menger determinant as
\be
V_{n_k}^2=\frac{(-1)^{n_k+1}}{2^{n_k}({n_k}!)^2}\det (\bm{A})\ .
\ee
The hypervolume is obviously zero for $n_k>4$ since all the $n_k$ vertices are located in three-dimensional space.  Thus, $\det(\bm{A})=0$ for $n_k>4$.

Note that, even for a general variogram, the Kriging system could also be degenerated in rare cases.
One could adjust $n_k$ or adopt the NP method instead for the grid points when this happens.

\bibliographystyle{apsrev4-1}
\bibliography{mybib150821}

\end{document}